\documentstyle[12pt]{article}                              
\def\word#1{\,\,\mbox{#1}\,\,}
\def\reff#1{(\ref{#1})}
\def\mass{e^{-\phi}{\nabla}^{2}\phi}
\def\nab{{\hat{\nabla}}^{2}\rho}
\def\beq{\begin{equation}}
\def\eeq#1{\label{#1}\end{equation}}

\def\rmunu{R_{\mu\nu}}
\def\dfrac#1#2{{\displaystyle\frac{#1}{#2}}}

\begin{document}
\begin{center}
{\bf Cosmological sector for localized mass and spin in 2+1 dimensional\\ 
topologically  massive gravity }
\end{center}
\medskip
\centerline {A. Edery and M. B. Paranjape}
\medskip
\begin{center}
{\it{Groupe de Physique des Particules, Laboratoire Ren\'ee J. A. Levesque,\\ 
Universit\'e de Montr\'eal, C.P. 6128, Montr\'eal ( Qu\'ebec),\\ Canada, H3C 3J7}}
\end{center}
\medskip
\centerline{Abstract}
\medskip
{\small The cosmological  sector to the full non-linear topologically massive
 gravity (TMG) is obtained for localized sources of mass  $m$ and
 spin $\sigma$ besides the asymptotically spinning conical flat sector
previously obtained \cite{Edery}. In a small region near but outside the sources, 
the metric resembles the spinning conical flat metric but we find that the mass $m$ creates a negative deficit angle of
$3m$ as opposed to $m$. Furthermore, it is not possible to
recover the results of pure Einstein gravity in the 
limit $\mu \to \infty$ unlike the flat sector. 
\\
\\
\centerline{I.\quad  Introduction}
\\
                                    
Gravity in 2+1 dimensions has attracted much interest in the last few years.
In supergravity theory, a mechanism in 2+1 dimensions originally proposed by
Witten \cite{Witten}, has been implemented as a possible solution to the
cosmological constant problem \cite{Strominger}. The authors demonstrated that
although unbroken supersymmetry prohibits the possibility of a cosmological
constant in the vacuum, massive states, more precisely massive soliton states,
cause the space-time to be conically flat and the generators of the
supersymmetry, which cause the usual spectrum doubling, fail to exist. That work
was done in the context of ordinary Einstein gravity (appropriately
supersymmetrized). It is well known that Einstein gravity is trivial in 2+1
dimensions i.e there is no propagating graviton and matter free regions 
are flat. 
However, 2+1 dimensions allows one to include the parity
violating gravitational Chern-Simons term \cite{Deser,Templeton} and the 
gravitational field 
now becomes a non-trivial propagating massive field. Such an inclusion is in
fact obligatory for a theory containing fermions and parity violation; it is
automatically induced via 1 loop quantum corrections with strength $N/2$ where
$N$ is the number of species of fermions \cite{Redlich}. The results of
\cite{Strominger} rested on the fact that exterior to
the soliton the space-time was conically flat. It 
becomes a compelling and interesting
exercise to discern what would the soliton and the exterior space-time 
represent for the case of topologically massive gravity (TMG). 
 
Many years ago, Vuorio \cite{Vuorio} obtained  
two solutions to TMG in vacuum ($T_{\mu\nu}=0$):
a trivial flat space
and a cosmological solution i.e. a homogenous space-time with constant curvature scalars. 
Localized sources can now be embedded in either of 
these two backgrounds. In the case of the flat 
background, solutions with localized mass $m$ and spin 
$\sigma$ 
have been obtained in the linearized \cite{Deser} and in the non-linear
theory \cite{Edery} and gives rise in both cases to an asymptotic spinning 
conical space-time. As mentioned in \cite{Deser}, the
cosmological solutions are not accessible in the linearized 
case and one is required to employ the full non-linear theory. Cosmological 
solutions have already been obtained in the non-linear theory using delta 
function
sources \cite{Clement}. It has been shown that delta function sources can be 
accomodated with torsion at the source \cite{Gerbert} but leads to 
inconsistencies in the torsion-free TMG theory
(see \cite{Edery}). An alternative and more consistent approach is to use 
non-singular sources that are arbitrarily localized.

We demonstrate that non-singular localized mass and spin sources in the full 
non-linear
TMG theory support a cosmological
sector. In this sector, the spin
source dominates the helical time structure of the metric at short distances outside the source
and the mass creates a negative deficit angle of $3m$ instead of the value $m$
found in the flat sector. Far from the sources, the metric resumes the form 
of the cosmological homogenous space-time
found by Vuorio. The cosmological sector is a disjoint sector in that it is not
possibe to recover pure Einstein gravity in the limit $\mu\to\infty$ where one
would expect the topological term to vanish.

\bigskip 
\bigskip
\centerline{II.\quad Field Equations}
\bigskip

We begin our work by writing down the well known field equations for TMG 
with 
energy-momentum tensor $T_{\mu\nu}$.
The Einstein field equations including a topological mass term is given by \cite{Deser,Clement}
(in units where $8 \pi G = 1$) 
\beq
\rmunu \,- \dfrac{1}{2} g_{\mu\nu}R +\dfrac{1}{\mu}C_{\mu\nu} =-{\kappa^2}\, T_
{\mu\nu}
\eeq{ein} 
where $\rmunu$
is the Ricci tensor , $R\,\equiv\,R_{\mu\nu}g^{\mu\nu}$ is the curvature scalar, and $C_{\mu\nu}$ is the 3 dimensional Weyl-Cotton tensor. We simplify
the field equations \reff{ein} by choosing a rotationally symmetric, stationary metric. The
most general form for such a metric is given by \cite{Vuorio1}
\beq
ds^2 = q^{2}{\left(\, dt + \psi(r)\,d\theta\,\right) }^2 - e^{\phi(r)}\,\left(dr^{2}+ r^{2}\,d\theta^{2}\right) \hspace{0.5cm} i = 1,2 
\eeq{metric}
To recover Vuorio's solution in the region exterior to the source we set
$q=1$ as in his work \cite{Vuorio} ( we will also show that the metric with 
$q=1$ does support solutions with localized mass and  spin).  
The functions $\psi(r)$ and $\phi(r)$ completely determine the metric. 
The scalar twist $\rho(r)$ is given by \cite{Edery}
\beq
\rho = e^{-\phi}\,\dfrac{\psi^{'} (r)}{r} .
\eeq{rho2}
where $\psi^{'}(r)\equiv \dfrac{d\psi(r)}{dr}$. To avoid a delta function 
$\rho$, $\psi(r)$ is set to zero at $r=0$ (see \cite{Edery} for more details).   
The (0,0), (0,j) and (i,j)
component field equations \reff{ein} are respectively \cite{Edery}
\beq
\dfrac{3}{4}\rho^2 - \dfrac{1}{2}\mass + \dfrac{1}{\mu}\rho^3 - \dfrac{1}
{2\mu}\nab - \dfrac{1}{2\mu}\,\rho\,\mass = - T_{00}
\eeq{t00}
\beq
\dfrac{\epsilon^{jk}e^{-\phi}}{2}\partial_{k}\left(\rho + \dfrac{3}{2\mu}
\rho^2 - \dfrac{1}{2\mu}\mass\right) = -  T_{0}^{j}
\eeq{t0j}
\beq
-e^{-\phi}\delta^{ij}\left(-\dfrac{\rho^2}{4}- \dfrac{1}{2\mu}\rho^3 +
 \dfrac{1}{2\mu}\nab + \dfrac{1}{4\mu}\,\rho\,\mass\right)- \dfrac{1}{2\mu}\hat{\nabla}^{i}\hat{\nabla}^{j}\,\rho = - T^{ij}\,.
\eeq{tij}
   
As in \cite{Edery}, we approach the problem of localized sources not by actually
specifying $T_{\mu\nu}$ but by examining the metric dependent side of the
field equations and drawing conclusions on the scalar twist $\rho$ and the
function $\phi$ if  $T_{\mu\nu}$ were localized.

\bigskip
\bigskip
\centerline{III.\quad Solving the Field Equations: Cosmological Sector}
\bigskip

When solving the field equations in the
vacuum i.e. $T_{\mu\nu}=0$, Vuorio obtains that the scalar twist $\rho(r)$ 
is a constant that
can have two possible values (corresponding to two metrics): one is given by 
$\rho(r)=0$ and $\mass =0$ which describes 
a flat metric ( Minkowski space) and the second is $\rho(r)=-2\mu/3$ and 
$\mass=2\mu^{2}/9$ which describes a cosmological space-time. 
To include sources of mass $m$ and spin $\sigma$ in this cosmological space, we introduce two 
functions $M(r)$ and $S(r)$ such that 
\beq
\mass=2\mu^{2}/9 + M(r)\quad \word{and}\quad \rho(r)=\dfrac{-2\mu}{3} +S(r)
\eeq{MS}
where $M(r)$ and $S(r)$ are zero outside the sources.
We will now see that $M(r)$ is proportional to the 
mass
density and $S(r)$ to the spin source density. 

When substituting $\rho$ above 
into the $T_{0j}$ ``spin" equation \reff{t0j} one observes that 
$\rho +3\rho^{2}/2\mu=
-S+3S^{2}/2\mu $. As expected the constant part of $\rho(r)$ i.e. $-2\mu/3$, 
has disappeared and does not contribute to the spin source. Clearly, $\rho(r)$ has
been replaced by $-S(r)$ in Eq. \reff{t0j}. In flat space, $\rho(r)$  was
the total conserved spin
source density (see \cite{Edery}) and we see that in the cosmological case, 
it is $-S(r)$ that takes on that role. 
  
The mass $m$ is defined as the total energy (i.e. the volume integral of 
$T_{00}$) when the spin source density  
$S(r)$ is zero. It will be localized in a region
from
$r=0$ to $r=\epsilon$ so that $M(r)$ and  $T_{00}$ are zero for $r>\epsilon$.
Substituting the quantities in Eq. \reff{MS} into
the $T_{00}$ equation \reff{t00}, one obtains         
\begin{eqnarray}
m &=&  \dfrac{1}{6}\int M(r) e^{\phi} \,d^2 r\nonumber\\
& =&  \dfrac{\pi}{3}\int_0^\epsilon \left(\nabla^2\phi -       
\dfrac{2\mu^{2}}{9} e^{\phi}\right) \, r\,dr \nonumber\\  
&=& \dfrac{\pi}{3}\,\left. r\,\phi^{'}\right|_0^\epsilon -
\dfrac {2\pi\mu^{2}}{27}\int_0^\epsilon e^{\phi} \, r\,dr.
\label{mass}
\end{eqnarray}
There are two terms to evaluate above. In the first term, the limit 
$r\to\epsilon$ can be obtained by matching the exterior to the interior solution at $r=\epsilon$. Outside the
mass source ($r>\epsilon$), $\mass = 2\mu^{2}/9$ and the most general
solution is given by \cite{Vuorio}
\beq  
\phi(r) =
2\ln\left(\dfrac{6n\,r^{n-1}}{\mu\,r_{0}^{n}\left[1-\left(\dfrac{r}{r_{0}}
\right)^{2n}\right]}
\right) \word{   for} r >\epsilon
\eeq{exterior}
where $n$ and $r_{0}$ are arbitrary constants and $r< r_{0}$. Note that 
$\phi(r)$ is invariant under the transformation $n\to -n$ so that 
positive and negative values for $n$ are both valid (nonetheless, this
invariance will have no direct physical consequences and for the sake
of clarity and  without loss of generality
we can assume $n$ is positive). With $\phi$ given
above, one obtains  
\beq
r\phi^{'} = 2(n-1) + \dfrac{4n(r/r_{0})^{2n}}{\left[1-(r/r_{0})^{2n}\right]}
\word{  for} r>\epsilon\\
\eeq{plus}
and
\beq
\lim_{r\to \epsilon}(r\phi{'})= 2(n-1)
\eeq{out}
where $\epsilon$ is sufficiently small so that terms proportional to 
$\epsilon$ are negligible and have been dropped in the above limit. 
For reasons given in the introduction, we do not allow delta function sources
and therefore we exclude the possibility that $\phi(r)\propto\ln r$ as $r\to
0$. Therefore
\beq
\lim_{r\to 0}(r\phi^{'}) = 0
\eeq{interior}
i.e. if $\lim_{r\to 0}(r\phi^{'}) =k$ where $k\ne 0$, then $\phi(r) \propto
k\ln r$ as $r\to 0$. With the above limits, Eqs. \reff{out} and \reff{interior}, the first term in Eq. \reff{mass} is   
\beq
\left. r\phi^{'}\right|_0^\epsilon = 2(n-1). 
\eeq{phiphi}
The mass $m$ is positive and therefore $n>1$. To evaluate the second
term in Eq. \reff{mass}, the behaviour of $\phi(r)$ in the source region needs 
to be known. In the exterior, $\phi(r)$ is negative and decreases as
$r\to\epsilon$ from the right i.e. behaves as $\ln r$ as $r\to\epsilon$. For
an elementary localized particle the mass density $M(r)$ (and hence
$\nabla^{2}\phi$) should be positive in the source region and this implies that
$\phi(r$) must continue to decrease
from $r=\epsilon$ to $r=0$.  It follows that the quantity $e^{\phi}$ is of the 
order $\epsilon$ (or smaller) in the source region
and that the second term in Eq. \reff{mass} can be neglected for $\epsilon$
sufficiently small.   
Substituting Eq. \reff{phiphi} into Eq. \reff{mass} one obtains
\beq
\left. r\phi^{'}\right|_0^\epsilon =\dfrac{3m}{\pi}\quad \word{and}\quad
n=1+\dfrac{3m}{2\pi}
\eeq{n}

We now turn to spin. The spin source $S(r)$ will be localized in a small region 
from $r=0$ to $r_{s}$ where $r_{0}\gg r_{s}\gg\epsilon$. 
By integrating the 
$T_0^j$ equation \reff{t0j}with $\rho(r)$ given by Eq. \reff{MS}
one obtains
\beq
 \int \epsilon^{ij} \, x^i \, \left( - T_{0}^{j}\right) e^{2\phi} \, d^2 r 
=-\pi\int_0^{r_{s}}  \left(-S(r) + \dfrac{3}{2\mu} S^{2}(r) 
 -  \dfrac{1}{2\mu}
\,\mass \right)^{\prime} \, e^{\phi}\,r^{2} \,dr \,.
\eeq{sigma}  
The integral of the third term                            can be readily evaluated using Eqs. \reff{plus} and \reff{interior} and yields
$-\left(3m/\mu\right)\left(1+3m/4\pi\right)$ (where the second term in
Eq. \reff{plus}, proportional to $(r_{s}/r_{0})^{2n}$, is negligible and has been dropped). For the integral of the  
first two terms we take  
$S(r)$ to be a
rapidly decreasing function
of $r$  with the condition that
$\lim_{r\to r_{s}} S(r) r^2 e^{\phi} = \lim_{r\to r_{s}}S^{2}(r)r^2
e^{\phi} =0$.
These integrals are well defined for any regular mass
distribution $\nabla^{2}\phi$, but depend on the actual profile and there will only be small differences for any two well localized mass distributions.
However, in the point mass limit
the result is
\beq
 \left( 2\pi+3m \right) \int_0^{r_{s}}\left( -S + \dfrac{3}{2\mu} S^2
\right) e^{\phi} r dr\,.
\eeq{parts}
Equation \reff{sigma} can now be expressed as  
\beq 
J_{S}\equiv 2\pi\int_0^{r_{s}}-S(r) e^{\phi} \,r\,dr = \sigma +\,  
\dfrac{3m}{\mu}\left(\dfrac{4\pi+3m}{4\pi+6m}
\right) -2\pi\int_0^{r_{s}} \dfrac{3}{2\mu} S^{2} e^{\phi} r dr \,.
\eeq{J}
where
\beq
\sigma \equiv \dfrac{2\pi}{2\pi+3m}\int \epsilon^{ij} \, x^i \, 
\left( - T_{0}^{j}\right) e^{2\phi} \, d^2 r .
\eeq{sigfig} 
Here $\sigma$ is identified as the bare spin i.e. 
it is equal to the spin source $J_{S}$ when the topological term is absent 
($\mu\rightarrow\infty$). The other two terms in Eq. \reff{J} represent the induced
spin (see \cite{Deser,Edery}).  

We now proceed to find the metric Eq. \reff{metric}. The function
$\psi(r)$, given by Eq. \reff{rho2} is
\begin{eqnarray} 
\psi(r)&=&\int_0^{r} \rho(r) e^{\phi} r dr\nonumber \\&=&\int_0^{r} S(r)
 e^{\phi} r
dr-\dfrac{2\mu}{3}\int_{0}^{r} e^{\phi} r dr\nonumber\\&=&
\dfrac{-J_{S}}{2\pi} -\dfrac{12n\, (r/r_{0})^{2n}}{\mu\, \left[1-(r/r_{0})^{2n}\right]}
\quad \word{for} r>r_{s}
\label{phimet}
\end{eqnarray}
where we see that there are two distinct contributions to $\psi(r)$: one from the spin source density $S(r)$ and one from the ``background" spin density $-2\mu/3$. In Eq. \reff{phimet}, the expression for $\phi$ exterior to the source
Eq. \reff{exterior}, was used to evaluate the second integral. For $r<r_{s}$, 
$\psi(r)$ and hence the metric, are well behaved but depend intimately on the
distribution of the mass and spin sources. Also, as mentioned in section II, $\psi(r)$ is zero at $r=0$ and the metric is therefore nonsingular.   
 
As in Vuorio's work \cite{Vuorio} we define new variables
\beq
\tilde{\theta}=n\theta, \,\,\,\sinh
x=\dfrac{2(r/r_{0})^n}{1-(r/r_{0})^{2n}} \quad\word{where}
n=1+\dfrac{3m}{2\pi}
\eeq{variable}
so that the metric, in terms of these new variables (dropping the tilde), is given by
\beq
ds^{2}=\left[dt -\left(\dfrac{J_{S}}{(2\pi+3m)}+\dfrac{6}{\mu}(\cosh x
-1)\right)d\theta\right]^{2} - \dfrac{9}{\mu^{2}}\left(dx^2 + \sinh^{2}x d\theta^{2}\right).
\eeq{finish}
This metric is similar to Vuorio's ``cosmological" metric \cite{Vuorio} but differs from it 
in 
two ways: the constant $n$, which appears in the redefinition of $\theta$, is 
not equal to 1 as in
Vuorio's case and our metric has a non-zero spin $J_{S}$. 

Since $n$ is not 1, our metric has a deficit angle i.e. with $n=1+3m/2\pi$ the new angle
$\theta$ runs from $0$ to $2\pi +3m$ instead of $2\pi$. Locally, this
represents a conical space with negative angular defect of $3m$. In the case 
of the flat background the deficit angle was simply the mass $m$ 
\cite{Deser,Edery}. The extra factor of three in the cosmological case is due 
to the spin-spin coupling term $\rho\,\mass/2\mu$ appearing in the $T_{00}$
equation \reff{t00}. This term contributes to the mass i.e. it is equal to 
$-\mass/3$ when $m$ is being defined, that is when $S(r)=0$ and 
$\rho$ is $-2\mu/3$. The 
coupling is therefore between a ``background" spin $\rho=-2\mu/3$ and an 
``induced" spin $\mass/2\mu$. In the flat
case, the spin-spin term makes no contribution to the mass because there is no
background spin. It is important to note that the part of the spin-spin term
which survives in equation \reff{t00}, $-M(r)/3$ is
independent of $\mu$ and therefore it is impossible to make it vanish in the
non-topological limit $\mu\to\infty$. 
Hence, the cosmological sector is a
disjoint sector and one cannot recover pure Einstein gravity
in the appropriate limit.

To see the effect of the spin $J_{S}$, note that  
in a region close to the source where $x$ is small, the term with 
$\cosh x -1$ in the metric is negligible
compared to $J_{S}$ and $\sinh^{2}x \approx x^2 $.  In the neighbourhood of 
the source the metric is 
\beq
ds^{2}=\left[dt - \dfrac{J_{S}}{(2\pi+3m)}\,d\theta\right]^{2} 
- \dfrac{9}{\mu^{2}}\left(dx^2 + x^{2} d\theta^{2}\right)
\eeq{finish2}
which is a spinning conical space with the spin $J_{S}$ governing the 
helical- time 
structure of the metric(see \cite{Jackiw,Vuorio} for a discussion on 
helical- time structure). 
Far from the sources, the metric \reff{finish} of course  describes the same 
space-time as Vuorio's i.e. a homogenous space-time with 
constant curvature scalars.

In conclusion, we have shown that there exists a distinct cosmologicial
sector for the exterior space-time to localized spin and mass sources in 2+1
dimensional topologically massive gravity in addition to the usual flat,
conical solution.  This is in contra-distinction to ordinary Einstein gravity 
in 2+1 dimensions and also in 3+1 dimensions which admit only a unique exterior
space-time. The ramifications of the existence of this 
sector should be investigated for the work of \cite{Strominger}, for example.

\end{document}